\documentclass[utf8,a4paper]{elsarticle}

\setcitestyle{square} 

\usepackage{url,hyperref,lineno,microtype,subcaption}
\usepackage{todonotes}
\usepackage[onehalfspacing]{setspace}
\usepackage{booktabs}
\usepackage{color}

\usepackage{amsmath,amssymb}

\begin{document}

\begin{frontmatter}

\title{Tensor Product Scheme for Computing Bound States of the Quantum Mechanical Three-Body Problem}

\address[tudelft]{Delft Institute of Applied Mathematics, Faculty of Electrical Engineering, Mathematics and Computer Science, Delft University of Technology Delft, The Netherlands}
\address[dlr-sc]{Department of High Performance Computing, Institute for Software Technology, German Aerospace Center (DLR), Cologne, Germany}
\address[dlr-qt]{Department of Theoretical Quantum Physics, Institute of Quantum Technologies, German Aerospace Center (DLR), Ulm, Germany}

\author[tudelft]{Jonas Thies}
\author[dlr-sc]{Moritz Travis Hof}
\author[dlr-qt]{Matthias Zimmermann}
\author[dlr-qt]{Maxim Efremov}

\begin{abstract}
We develop a computationally and numerically efficient method to calculate binding energies and corresponding wave functions of quantum mechanical three-body problems in low dimensions. Our approach exploits the tensor structure of the multidimensional stationary Schr\"odinger equation, being expressed as a discretized linear eigenvalue problem. In one spatial dimension, we solve the three-body problem with the help of iterative methods. Here the application of the Hamiltonian operator is represented by dense matrix-matrix products. In combination with a newly-designed preconditioner for the Jacobi-Davidson QR, our highly accurate tensor method offers a significantly faster computation of three-body energies and bound states than other existing approaches. For the two-dimensional case, we additionally make use of a hybrid distributed/shared memory parallel implementation to calculate the corresponding three-body energies. Our novel method is of high relevance for the analysis of few-body systems and their universal behavior, which is only governed by the particle masses, overall symmetries, and the spatial dimensionality. Our results have straightforward applications for ultracold atomic gases that are widespread and nowadays utilized in quantum sensors.

\end{abstract}

\begin{keyword}
 Schr\"odinger equation \sep three-body problem \sep pseudospectral method \sep tensor product structure \sep Jacobi-Davidson method
\end{keyword}

\end{frontmatter}

\section{Introduction}
\label{sec:introduction}

The quantum mechanical few-body problem is of particular interest for the physics community.
On the one hand, it determines the features of interacting nuclei, atoms, or molecules as bodies living on very different length scales. On the other hand, in certain regimes these systems display a universal behavior that is independent of the details of the interaction between the particles, but governed by the particle masses, overall symmetries, as well as the dimensionality of space.
The complexity and beauty of this problem has motivated numerous researchers to explore these systems by using theoretical, numerical, and experimental approaches.

An outstanding example for the above mentioned type of universality is the Efimov effect~\cite{Efimov_PLB_1970,Efimov_NPA_1973}, describing the emergence of an infinite sequence of universal states of three bosonic particles with $s$-wave resonant pair interactions in three dimensions. Lower dimensional systems, such as three fermionic particles confined to two dimensions, also display surprising universal phenomena like the so-called ``super Efimov effect''~\cite{Nishida_PRL_2013,Moroz_PRA_2014,Gridnev_JPA_2014,Volosniev_JPB_2014}. 

In addition, also mass-imbalanced three-body systems can be governed by universal features. Recently, it has been demonstrated that a heavy–heavy–light system confined to one dimension (1D) displays universality not only in the discrete spectrum~\cite{Happ2019threebody,Happ2021proof}, but also in the continuum ~\cite{Happ2021submitted}. Here universal three-body energies and wave functions emerge once the heavy-light interactions are tuned towards the ground- or exited-state threshold, respectively, that is the binding energy of the ground or exited state in the heavy-light system approaches zero. In this limit, the three-body binding energies and wave functions for arbitrary short-range heavy-light interactions converge to the respective ones for the zero-range interaction.

In order to provide an accurate description of the universal behavior in these and higher-dimensional systems, novel analytical and numerical tools are required. For instance, in Ref.~\cite{Happ2019threebody} three-body energies and the corresponding wave functions of the bound states are computed with the pseudo-spectral method~\cite{boyd2000chebyshev,Trefethen}, where the Hamiltonian is represented by a sparse matrix. Then the Krylov subspace method is applied to determine lowest eigenvalues and the corresponding eigenvectors. However, with an increasing number of grid points in each dimension, the matrices and vectors grow rapidly: the three-body problem in $d$ space dimensions yields a $2d$-dimensional linear eigenvalue problem after removing the center-of-mass degree of freedom. When discretized with $n$ grid points in each direction, a single vector representing the three-body wave function has the size $n^{2d}$. The pseudo-spectral discretization used in~\cite{Happ2019threebody} also leads to $\mathcal{O}\left(n^{2d-1}\right)$ dense blocks in the sparse matrix representation of the Hamiltonian, each of size $n\times n$. Thus, such an approach is severely limited by the `curse of dimensionality'. 

In this article, we present a novel computational approach to analyze  three-body problems with local two-body interactions in 1D and 2D. In particular, we exploit the tensor product structure of the problem and avoid to store redundant blocks of the matrix. In this way, we achieve a very high computational efficiency. In order to accelerate the convergence compared to the Krylov method used in Ref.~\cite{Happ2019threebody}, we utilize the Jacobi-Davidson iteration scheme and introduce a preconditioner for the 1D three-body system. 
By extending our approach to the 2D case, we show for the first time the universal behavior of the heavy-heavy-light three-body system when the ground-state energy of the heavy-light subsystems approaches zero. The methods and results presented in this article constitute the first steps towards the ab-initio simulation of quantum systems in 2D and 3D with a larger number of particles involved.

Our article is structured as follows. In Section~\ref{sec:few-body_systems} we introduce quantum mechanical few-body systems in 1D as well as 2D and present the corresponding eigenvalue equations determining the energies and stationary wave functions. In Section~\ref{sec:tensor_form} we describe the discretization scheme and explain how it naturally enables a tensor formulation of the Hamiltonian operator. The Jacobi-Davidson iteration scheme is revisited in Section~\ref{sec:eigensolver}. In addition, the implementation of the discretized operators in 1D and 2D are discussed in terms of hardware efficiency. For the 1D case, we devise a novel preconditioning technique to accelerate the convergence of the Jacobi-Davidson method, and successfully determine the eigenpairs corresponding to the three-body bound states. Numerical results presented in Section~\ref{sec:results} show the superior performance and convergence properties of our method in 1D and 2D. Moreover, we show the universal behavior of the three-body system in 2D. Finally,  we provide in Section~\ref{sec:conclusion} concluding remarks and indicate directions of future research.
 

\section{Few-body systems}
\label{sec:few-body_systems}

First, we consider a system composed of two interacting particles, a heavy one of mass $M$ and a light one of mass $m$. In dimensionless units, the relative motion of these quantum particles is governed by the stationary Schr\"odinger equation
\begin{equation}
\left[-\frac{1}{2}\Delta_{\vec{\xi}}-v_0f\left(\xi\right)\right]\psi^{(2)}(\vec{\xi})=\mathcal{E}^{(2)}\psi^{(2)}(\vec{\xi})
\label{eq:Schroedinger_equation_two_body}
\end{equation}
for the wave function $\psi^{(2)}(\vec{\xi})$ and two-particle energy $\mathcal{E}^{(2)}$, where $\Delta_{\vec{\xi}}$ denotes the Laplace operator with respect to the relative coordinate $\vec{\xi}$. Here, we have assumed that the interaction of the two particles is described by an attractive potential $-v_0 f(\xi)$ of magnitude $v_0>0$ and shape $f(\xi)$ as a function of the relative distance $\xi\equiv \left|\vec{\xi}\ \right|$. 

\begin{figure}
\begin{center}
       \includegraphics[width=0.4\textwidth]{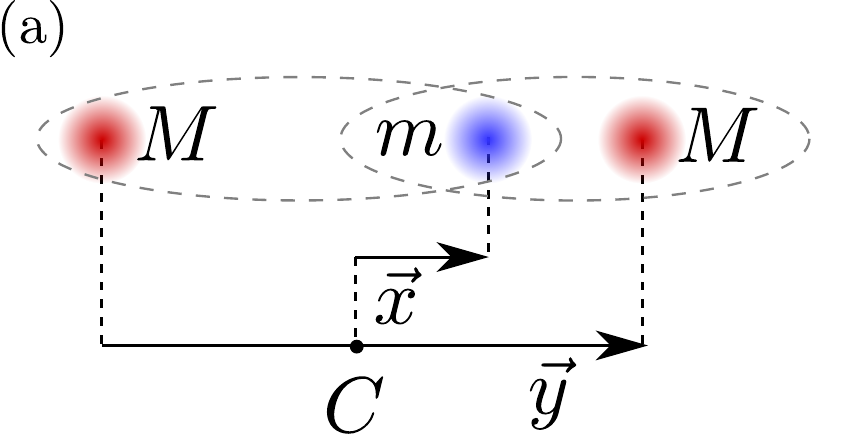}
        \hspace*{1cm}
         \includegraphics[width=0.4\textwidth]{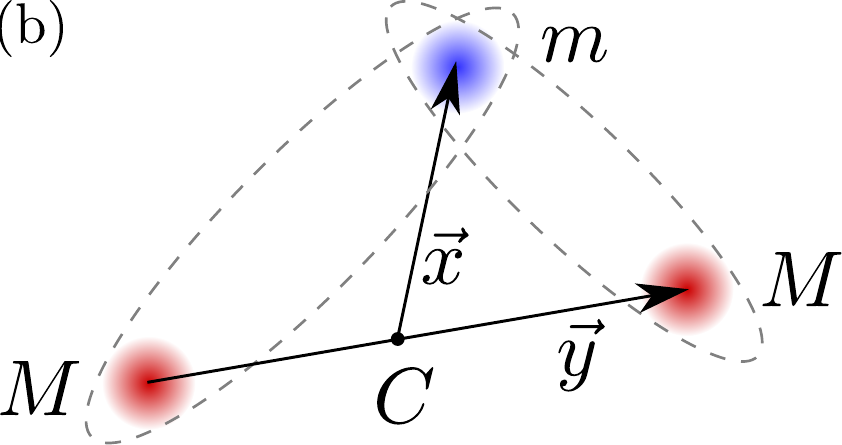}
\end{center}
    \caption{Three-body system consisting of two heavy particles of mass $M$ and a light one of mass $m$  confined to (a) one or (b) two spatial dimensions. We only allow for interactions between heavy and light particles, as indicated by the gray ellipses, and describe the system in terms of the Jacobi coordinates $\vec{x}$ and $\vec{y}$, where $C$ denotes the center-of-mass of the two heavy particles.
    \label{fig:3body_scheme}
    }
\end{figure}

Next, we turn to the mass-imbalanced three-body system displayed in Fig.~\ref{fig:3body_scheme} and confined to (a) one or (b) two spatial dimensions. This system is described by the dimensionless form of the stationary Schr\"odinger equation
\begin{equation}
\left[-\frac{\alpha_x}{2}\Delta_{\vec{x}}-\frac{\alpha_y}{2}\Delta_{\vec{y}}+V(\vec{x},\vec{y})\right]\psi=\mathcal{E}\psi
\label{eq:Schroedinger_equation_three_body}
\end{equation}
for the three-particle wave function $\psi=\psi\left(\vec{x},\vec{y}\right)$ corresponding to the three-particle energy $\mathcal{E}$, as introduced in more detail in Refs.~\cite{Happ2019threebody,Happ2021submitted} for the one-dimensional case. Here $\Delta_{\vec{x}}$ and $\Delta_{\vec{y}}$ denote the Laplace operator with respect to the relative coordinate vectors $\vec{x}$ and $\vec{y}$, respectively. The positive coefficients $\alpha_x= 2/(1+\alpha)$ and $\alpha_y=(1+2\alpha)/(2+2\alpha)$ are determined by the mass ratio $\alpha\equiv M/m$ of the heavy and light particle.

For the case of non-interacting heavy particles, the interaction term $V(\vec{x},\vec{y})$ in Eq.~\eqref{eq:Schroedinger_equation_three_body} reads
\begin{equation}
V(\vec{x},\vec{y})=-v_0 f\left(\left|\vec{x}+\frac{1}{2}\vec{y}\right|\right)-v_0 f\left(\left|\vec{x}-\frac{1}{2}\vec{y}\right|\right), 
\label{eq:definition_potential}
\end{equation}
where $-v_0 f(\xi)$ models the interaction potential between the light particle and each heavy one. Here $\left|\vec{x}\pm \vec{y}/2\right|$ is the respective relative distance in (a) one or (b) two spatial dimensions, as shown in Fig.~\ref{fig:3body_scheme}. 

Our scheme to solve the three-body problem, Eq.~\eqref{eq:Schroedinger_equation_three_body}, consists of several steps. First, we choose a particular binding energy  $\mathcal{E}_0^{(2)}$ for the two-body system of the heavy and light particle. Then we determine the corresponding depth $v_0$ of the potential $-v_0 f(\xi)$ such that the two-body Schr\"odinger equation \eqref{eq:Schroedinger_equation_two_body} has the ground state solution with  energy $\mathcal{E}^{(2)}=\mathcal{E}^{(2)}_0$.  Next, we solve the three-body Schr\"odinger equation \eqref{eq:Schroedinger_equation_three_body} for this particular potential depth $v_0$ and select solutions with an energy $\mathcal{E}$ smaller than the two-body threshold given by $\mathcal{E}^{(2)}_0$. In this way, we determine the wave functions $\psi\left(\vec{x},\vec{y}\right)$ and corresponding energies $\mathcal{E}$ of the {\it three-body bound states} associated with this particular two-body interaction. 

In physics, most two-body potentials vanish either exponentially or polynomially as $\xi\rightarrow \infty$. In the present article, we consider both cases and focus on an attractive  potential of Gaussian shape 
\begin{equation}
f_\mathrm{G}(\xi)=\exp\left(-\xi^2\right)
\label{eq:Gaussian_shaped_potential}
\end{equation}
and a potential whose shape
\begin{equation}
f_\mathrm{L}(\xi)=\frac{1}{(1+\xi^2)^3}
\label{eq:Lorentzian_shaped_potential}
\end{equation}
is determined by the cube of a Lorentzian. However, we emphasize that our approach is also valid for potentials that feature  a different shape as a function of the relative coordinate $\vec{\xi}$.

\section{Discretization and tensor formulation}
\label{sec:tensor_form}

In this section, we briefly describe the discretization of the three-body Schr\"odinger equation by pseudo-spectral methods and, in particular, the Lagrange-mesh method \cite{BAYE2015Lagrange-mesh}.   In contrast to finite difference or finite element methods, pseudo-spectral methods lead to dense matrices for one-dimensional problems.  As a consequence, we obtain a faster convergence rate with respect to the number of grid points which is geometric on a finite domain and usually subgeometric on an infinite domain~\cite{boyd2000chebyshev}.

Since the three-body Hamiltonian is multi-dimensional, we aim to exploit its tensor structure when implementing the matrix-vector multiplication in an iterative eigenvalue solver. Our tensor method drastically reduces the memory requirement for the operator when solving higher-dimensional eigenvalue problems.

\subsection{Discretization}

We apply a pseudo-spectral method to build a matrix representation of the Schr\"odinger equation \eqref{eq:Schroedinger_equation_three_body} of the three-body problem.
For this purpose, we consider Chebyshev polynomials as basis functions on a finite domain in each dimension. The corresponding grid points are then associated with the roots of the Chebyshev polynomials. By using an algebraic map \cite{boyd2000chebyshev,boydMap}, we project these grid points to the infinite real domain.

After discretization, the Schr\"odinger equation  \eqref{eq:Schroedinger_equation_three_body} takes the form 
\begin{equation}
H \vec{\psi}=\mathcal{E}\vec{\psi},
\label{eq:matrix_evp}
\end{equation}
which is a linear eigenvalue problem for the matrix $H$ with eigenvalue $\mathcal{E}$ and eigenvector $\vec{\psi}$.

For the one-dimensional three-body problem, the eigenvector
\begin{equation}
    \vec{\psi}^{(\mathrm{1D})}\equiv \{\psi_{0,0}, \psi_{0,1},\ldots,\psi_{0, N_{y_1}-1}, 
    \psi_{1,0},\ldots,\psi_{N_{x_1}-1, N_{y_1}-1} \}^T
\end{equation}
corresponds to the wave function $\psi(\vec{x},\vec{y})=\psi^{(\mathrm{1D})}(x_1,y_1)$ in Eq.~\eqref{eq:Schroedinger_equation_three_body} evaluated at the grid points  $\left(x_1^{(i)},y_1^{(j)}\right)$, yielding the entries $\psi_{i,j}=\psi^{(\mathrm{1D})}(x_1^{(i)},y_1^{(j)})$ with $i=0,1,\ldots, N_{x_1}-1$ and $j=0,1,\ldots,N_{y_1}-1$. Here $N_{x_1}$ and $N_{y_1}$ denote the number of grid points in the respective direction.

The matrix $H$ in Eq.~\eqref{eq:matrix_evp} reads
\begin{equation}
H^{(\mathrm{1D})}=- \frac{\alpha_x}{2} \left(D_{x_1 x_1}\otimes \mathbb{I}_{y_1}\right) - \frac{\alpha_y}{2}\left(\mathbb{I}_{x_1} \otimes D_{y_1y_1}\right) +v_0(F_+ + F_-).
\label{eq:discretized_Schroedinger_equation_1D}
\end{equation}
Here $D_{x_1 x_1}$ and $D_{y_1 y_1}$ are dense (generally non-symmetric) matrices with sizes $N_{x_1}\times N_{x_1}$ and $N_{y_1}\times N_{y_1}$ corresponding to the partial second derivatives $\partial^2/\partial x_1^2$ and $\partial^2/\partial y_1^2$ from the Laplace operators $\Delta_{\vec{x}}$ and $\Delta_{\vec{y}}$ in Eq.~\eqref{eq:Schroedinger_equation_three_body}, respectively. Moreover, $\mathbb{I}_{x_1}$ and $\mathbb{I}_{y_1} $ denote the identity matrices of corresponding size.
In addition, the diagonal matrices $F_\pm$ result from evaluating the functions $f\left(\left|\vec{x}\pm\vec{y}/2\right|\right)$ in Eq.~\eqref{eq:definition_potential} at the grid points $x_1^{(i)}$ and $y_1^{(j)}$. More details on the discretization procedure and the exact form of the matrices in Eq.~\eqref{eq:discretized_Schroedinger_equation_1D} can be found in Appendix~B of Ref.~\cite{Happ2019threebody}.

Similarly, for the three-body problem in two dimension we perform a discretization of the wave function $\psi\left(\vec{x},\vec{y}\right)=\psi^{(\mathrm{2D})}(x_1,x_2,y_1,y_2)$ with respect to the grid points $(x_1^{(i)},x_2^{(j)},y_1^{(k)},y_2^{(l)})$. The matrix $H$ in Eq.~\eqref{eq:matrix_evp} then reads
\begin{equation}
\label{eq:Ham2D}
H^{(\mathrm{2D})}=-\frac{\alpha_x}{2}\left[\left(\boldsymbol{D}_{x_1x_1}+\boldsymbol{D}_{x_2x_2}\right)\otimes \mathbb{I}_y\right]-\frac{\alpha_y}{2}\left[\mathbb{I}_x\otimes (\boldsymbol{D}_{y_1y_1}+\boldsymbol{D}_{y_2y_2}) \right]+v_0\left(\boldsymbol{F}_+ + \boldsymbol{F}_-\right),
\end{equation}
where $\boldsymbol{D}_{x_1x_1}=(D_{x_1x_1} \otimes \mathbb{I}_{x_2})$, $\boldsymbol{D}_{x_2x_2}=(\mathbb{I}_{x_1} \otimes D_{x_2x_2})$, $\mathbb{I}_x=\mathbb{I}_{x_1}\otimes\mathbb{I}_{x_2}$, etc. Also here the diagonal matrices $\boldsymbol{F}_\pm$ result from evaluating the function $f\left(\left|\vec{x}\pm \vec{y}/2\right|\right)$ in Eq.~\eqref{eq:definition_potential} at the corresponding grid points.


\subsection{Operator application for the 1D case\label{sec:op1D}}

When using an iterative method for solving the linear eigenvalue problem given by 
the discretized Schr\"odinger equation \eqref{eq:matrix_evp}, only the effect of the linear operator on a given vector has to be implemented.
Our 1D Hamiltonian $H^{(\mathrm{1D})}$, Eq.~\eqref{eq:discretized_Schroedinger_equation_1D}, can be abstractly written in the form
\begin{align}
\label{eq:tensor_form}
T_{V,a_1,a_2} = & a_1\left(C_1\otimes\mathbb{I}_2\right)+a_2\left(\mathbb{I}_1\otimes C_2\right)+V
\end{align}
with $C_1=D_{x_1 x_1}$, $C_2=D_{y_1 y_1}$, and a sparse (in our case diagonal) matrix $V=v_0(F_+ + F_-)$.
Note that we do not assume any structure for the potential operator $V$, in particular, it does not have to be of tensor structure $\mathbb{I}_1\otimes A_1 + A_2 \otimes \mathbb{I}_2$ with matrices $A_1$ and $A_2$.

The application of the operator $T_{V,a_1,a_2}$, Eq.~\eqref{eq:tensor_form}, to a vector $w$ can be efficiently implemented by using dense matrix-matrix products. 
Let $C_1\in\mathbb{R}^{N_1\times N_1}$, $C_2\in\mathbb{R}^{N_2\times N_2}$, $w\in\mathbb{R}^{N_1N_2}$, and $W=\mathrm{reshape}(w,N_2,N_1)$ denote the interpretation of $w$ as an $N_2\times N_1$ matrix. Then we have
\begin{align}
\label{eq:tensor_op1D}
T_{V,a_1,a_2}\cdot w = \mathrm{reshape}\left(a_2 C_2 \cdot W + a_1 W \cdot C_1^T, N_1 N_2, 1\right) + V\cdot w,
\end{align}
where the $\mathrm{reshape}$ operation is used to interpret the resulting $N_2 \times N_1$-matrix as a vector of length $N_1N_2$. Here we emphasize that $\mathrm{reshape}$ does not incur any data movement, namely it is just a re-interpretation of a vector as a matrix stored in column-major ordering, and vice versa.

Assuming that $N_{x_1}=N_{y_1}=n$, the storage requirement of the Hamiltonian operator in 1D is now only $\mathcal{O}(n^2)$, as compared to $\mathcal{O}(n^3)$ when storing it in a sparse matrix format. 
The performance of the $\mathcal{O}(n^3)$ arithmetic operations is limited by the floating point units of the hardware (compute bound).
In general, an operation is compute bound if the arithmetic intensity $I_c$, defined as the ratio of required floating point operations (flops) and bytes of memory transferred, is larger than the machine balance $I_M$, defined as the ratio of the peak floating point performance and the memory bandwidth of the hardware.

For our matrix-matrix products, $I_c=\mathcal{O}(n)\ \mathrm{byte}/\mathrm{flops}$, which is above the machine balance $I_M$ on typical CPUs. 
For example, the Intel Xeon Gold 6248R CPU used for our 2D simulations in Section~\ref{sec:results2D} achieves a memory bandwidth (pure load) of 268 GB/s and can perform about 4\,600 Gflop/s when running at 3.0 GHz, yielding  $I_M\approx 17\ \mathrm{flops}/\mathrm{byte}$.
Applying the operator $V$ to $w$ is a memory-bound operation in general, because it requires only two flops per matrix entry loaded. Consequently, its cost is $\mathcal{O}(n^2)$ memory transfers.

Thus, if $H^{\mathrm{(1D)}}$ is represented as a sparse matrix, then loading and applying the operator $T_{V,a_1,a_2}$, Eq.~\eqref{eq:tensor_op1D}, would cost $\mathcal{O}(n^3)$ memory transfers and still $\mathcal{O}(n^3)$ flops.
The operation is then memory bound, as $I_c=\mathcal{O}(1)$.
So the high values of $I_M$ in modern HPC hardware, like CPUs and GPUs, can lead to a speed-up of about a factor 100 when going from the sparse matrix representation to the tensor operations. This is exactly a key idea in this article. As an example, for the processor mentioned above and $I_c=1$, we would achieve a performance 200 times below the peak floating point performance for double precision data.

\subsection{Operator application for the 2D case\label{sec:op2D}}

For the 2D problem, both the operator $H^{(\mathrm{2D})}$, Eq.~\eqref{eq:Ham2D}, and its components 
$\boldsymbol{D}_{xx}=\boldsymbol{D}_{x_1x_1}+\boldsymbol{D}_{x_2x_2}, \boldsymbol{D}_{yy}=\boldsymbol{D}_{y_1y_1}+\boldsymbol{D}_{y_2y_2}$ are of the form Eq.~\eqref{eq:tensor_form} with particular choices of $a_1$, $a_2$, and $V$.
Thus, we can now apply Eq.~\eqref{eq:tensor_op1D} in a nested way. Indeed, let
$W=\mathrm{reshape}(w,N_{y_1}N_{y_2},N_{x_1}N_{x_2})$ and $N=N_{x_1}N_{x_2}N_{y_1}N_{y_2}$, resulting in
\begin{align}
\label{eq:tensor_op2D}
H^\mathrm{(2D)}\cdot w = \mathrm{reshape}\left(-\frac{\alpha_y}{2}\boldsymbol{D}_{yy} \cdot W - \frac{\alpha_x}{2}W \cdot \boldsymbol{D}_{xx}^T, N,1\right) + V\cdot w.
\end{align}

If we assume that $N_{x_1}=N_{x_2}=N_{y_1}=N_{y_2}=n$, then loading the operator still requires $\mathcal{O}(n^2)$ memory transfers. However, the vector $w$ now actually represents 4D tensors and has a storage requirement of $\mathcal{O}(n^4)$. Hence, the total amount of memory transferred is $\mathcal{O}(n^4)$. 
For each of the $n^2$ columns of $W$, $\mathcal{O}(n^3)$ flops are performed, amounting to $\mathcal{O}(n^5)$ in total.
The computational intensity is therefore still $\mathcal{O}(n)$ and the operation is compute bound as before.

In practice, we implement the term $W \boldsymbol{D}_{xx}^T$ in Eq.~\eqref{eq:tensor_op2D} as $(\boldsymbol{D}_{xx} W^T)^T$, so that $n^4$ vector elements
have to be read and written twice in each operator application due to the transpose operations. In addition, we have also developed a distributed memory implementation where
the columns of $W$ and $W^T$ are distributed among several processes running on different nodes of a cluster. In this way, the transpose operations translate
to an `all to all' communication pattern. Whenever the communication for a single column is complete, the corresponding product with $\boldsymbol{D}_{xx}$ can be performed (see also Section~\ref{sec:results2D}). Overall this yields a very efficient and scalable implementation. However, we also remark that the overhead of transposing a tensor twice can be avoided by using an optimized implementation of a tensor contraction, such as GETT~\cite{paolo2018gett}.



\section{Jacobi-Davidson and preconditioning}\label{sec:eigensolver}

In this section, we briefly introduce the Jacobi-Davidson QR (JDQR) method \cite{jacdav}. Compared to the Arnoldi-type iteration (Krylov-Schur), which is implemented in MATLAB's \texttt{eigs} command, JDQR offers some flexibility when solving the so-called correction equation. We use this
flexibility to improve the convergence dramatically, by introducing a preconditioned iteration for the correction equation and exploiting again the tensor
structure of the matrices.

\subsection{The Jacobi-Davidson QR Method}

We use a Matlab implementation of the JDQR method \cite{jacdav} that is
suitable for computing a few exterior eigenvalues of a non-Hermitian matrix. 
The algorithm computes a partial QR decomposition of a matrix $A$ by applying a Newton process to the system of equations
\begin{align}\label{eq:jdqr}
  \begin{cases}AQ - QR &= 0,\\
              -\frac{1}{2}Q^TQ + \frac{1}{2}\mathbb{I} &= 0.
  \end{cases}
\end{align}

The Newton updates are utilized to extend the search space spanned by $Q$, whereas the standard Ritz pairs are used for approximating eigenpairs of the matrix $A$. Whenever the basis spanning the search space reaches a maximum size $m_\mathrm{max}$, it is compressed into $m_\mathrm{min}$ vectors by (implicitly) applying a truncated Singular Value Decomposition (SVD) to retain only the most relevant directions.

The Newton process requires solving the correction equation 
\begin{align}\label{eq:jdcor}
  (\mathbb{I}-\tilde Q \tilde Q^T)(A-\theta\mathbb{I})(\mathbb{I}-\tilde Q\tilde Q^T) \Delta q  &= -(A q - q \theta)
\end{align}
for the new basis vector $\Delta q$ in every outer iteration $i$. Here $\tilde Q$ contains the approximation
$q\approx Q_i$ for the current eigenvector and any previously converged (`locked') eigenvectors. Moreover, $\theta\approx R_{ii}$ denotes the current approximate eigenvalue.

The deflation operator $\mathbb{I}-\tilde Q \tilde Q^T$ improves the conditioning of the shifted matrix $A-\theta\mathbb{I}$.
We employ a Generalized Minimal Residual (GMRES) method to solve Eq.~\eqref{eq:jdcor} with additional acceleration resulting from preconditioning, as discussed in the next section. Further details on how the preconditioner is combined with the projections can be found in Ref. \cite{Thies2020phist}.

\subsection{Preconditioning\label{sec:precon}}

In order to improve the convergence of the GMRES correction solver, we introduce a
shifted version of the Hamiltonian that ignores the potential $V$. Moreover, we neglect $V$ in the
preconditioner allowing us to exploit the tensor product structure of the differential
operator, even if $V$ does not have tensor structure, as discussed above.

In the 1D case and for small values of the potential depth $v_0$, the matrix $H^{\mathrm{(1D)}}$, Eq.~\eqref{eq:discretized_Schroedinger_equation_1D}, can be approximated by an operator of the form $T_{0,1,1}$ given by Eq.~\eqref{eq:tensor_op1D}, where $C_{1}=-\frac{\alpha_{y}}{2}D_{y_1 y_1}, C_2=-\frac{\alpha_x}{2}D_{x_1 x_1}$.
For some scalar $\sigma$, linear system with $T_{0,1,1}-\sigma\mathbb{I}$ and some right-hand side $b$, respectively $B=\mathrm{reshape}(b,N_{y_1},N_{x_1})$, can be solved for $w$ with the help of the Sylvester equation
\begin{align}
\left(C_1-\sigma_1\mathbb{I}_1\right) W + W \left(C_2-\sigma_2\mathbb{I}_2\right)^T = B
\label{eq:sylvester_equation}
\end{align}
with $\sigma=\sigma_1+\sigma_2$. 

For our system, the shift  \mbox{$\sigma_{1,2}=-\alpha_{y,x}\mathcal{E}_0^{(2)}/|\alpha_x+\alpha_y|$} is a good choice. In this way, the preconditioner approximates the shift-invert operator near the value $-\mathcal{E}^{(2)}_0$ of the two-body binding energy, which is close to the desired eigenvalues.

Bartels and Steward~\cite{barstew} have introduced a direct method for solving the  Sylvester equation~\eqref{eq:sylvester_equation}. It requires a Schur decomposition of the shifted matrices $C_{1,2}-\sigma_{1,2}\mathbb{I}$, a
combination of two dense matrix-matrix products, and a special forward/backward substitution
with the Schur factors. Since the matrices involved remain the same throughout the JDQR process, the Schur factorization has to be performed only
once. Applying our preconditioner again has a computational cost of $\mathcal{O}(n^3)$ and requires $\mathcal{O}(n^2)$ data transfers, so that the performance characteristics of the overall algorithm are unchanged.


Unfortunately, we cannot straightforwardly extend our preconditioner, which is a direct solver for a shifted operator, to the 2D case. Indeed, the Schur decomposition of the operators $\boldsymbol{D}_{xx}$ and  $\boldsymbol{D}_{yy}$ cannot be simply represented  as a sum of
Kronecker product terms. Instead one might use an iterative procedure to approximate the effect of the shift-invert operator in 2D and we do not consider such techniques here. An an alternative approach, we focus on a distributed memory implementation of the 2D operator application, in order to accommodate the significant memory requirement for storing the vectors, which represent 4D tensors. Numerical results for the 2D case without preconditioning are shown in Section~\ref{sec:results2D}.

\section{Numerical and performance results}\label{sec:results}
In this section, we present the results for our numerical study of the quantum mechanical three-body problem in one and two spatial dimensions.
We investigate the convergence of the three-body energies as a function of the number of grid points. Moreover, we compare the performance of three iterative eigenvalue solvers: Krylov-Schur, Jacobi-Davidson QR with and without preconditioning. 

\subsection{Results for the three-body problem in 1D}
\label{sec:Results_1D}
We are now in the position to compute the bound states of the 1D three-body system introduced in Section~\ref{sec:few-body_systems} and depicted in Fig.~\ref{fig:3body_scheme} (a). We first determine the potential depth $v_0$ for the Gaussian-shaped potential $f_\mathrm{G}$, Eq.~\eqref{eq:Gaussian_shaped_potential}, such that it corresponds to a specific two-body binding energy $\mathcal{E}_0^{(2)}$. 
For this purpose, we choose a particular value of the two-body binding energy $\mathcal{E}^{(2)}=\mathcal{E}_0^{(2)}$ in the discretized version of the Schr\"odinger equation~\eqref{eq:Schroedinger_equation_two_body} in 1D and solve the generalized eigenvalue problem for the lowest eigenvalue $v_0$.
For the two-body binding energies $\mathcal{E}_0^{(2)}=10^{-1},10^{-2},$ and $10^{-3}$, the corresponding potential depths $v_0$ are listed in Table~\ref{tab:2b_binding_energies}. 

Next, we use these parameters to solve the discretized Schr\"odinger equation~\eqref{eq:matrix_evp} for the three-body problem in 1D with the Hamiltonian matrix given by Eq.~\eqref{eq:discretized_Schroedinger_equation_1D}. 
To increase the accuracy of our method for a given grid resolution, we apply the parity selection rule, reducing the problem size by a factor of $2^2$ for a requested accuracy. More precisely, we exploit the symmetry properties of the basis functions used for discretization, as outlined in Ref.~\cite{Trefethen}. Consequently, bosonic and fermionic bound states have to be computed separately. These particular states
are characterized by even respectively odd wave functions with regard to the transformation $\vec{y}\to-\vec{y}$, corresponding to the exchange of the two heavy particles, see Fig.~\ref{fig:3body_scheme}.

For bosonic and fermionic heavy particles and a mass ratio $\alpha=M/m=20$ of heavy and light particles, we list the resulting ratios $\mathcal{E}/\mathcal{E}_0^{(2)}$ of three-body and two-body binding energy in Table~\ref{tab:2b_binding_energies}. Our results coincide with the ones presented in Ref.~\cite{Happ2019threebody} for the Gaussian-shaped interaction potential $f_\mathrm{G}$, Eq.~\eqref{eq:Gaussian_shaped_potential}. Moreover, as $\mathcal{E}_0^{(2)}\to 0$ these ratios approach the universal values listed in Table~1 of Ref.~\cite{Happ2021submitted}. 

\begin{table}
\caption{\label{tab:2b_binding_energies}
Computed ratio $\mathcal{E}/\mathcal{E}_0^{(2)}$ of three-body and two-body binding energies in 1D for the case of two heavy bosons or fermions as obtained by solving the discretized Schr\"odinger equation~\eqref{eq:matrix_evp}. The calculations are performed for the mass ratio $M/m=20$ of heavy and light particles, interacting via a Gaussian shaped potential $f_\mathrm{G}$, Eq.~\eqref{eq:Gaussian_shaped_potential}. Here the potential depth $v_0$ has been chosen such that it corresponds to a particular two-body binding energy  $\mathcal{E}_0^{(2)}$.}
\centering
{
\begin{tabular}{cccc}
\toprule
   $\mathcal{E}_0^{(2)}$ & $v_0$  in 1D  & bosons & fermions    \\
\midrule  
 $10^{-1}$        &  0.34459535    & -2.47603458  & -1.82589653\\   
                  &                 & -1.41279329  & -1.18259157\\
                  &                 & -1.06093864  & -1.02845702\\
                  \midrule
 $10^{-2}$        &  0.08887372    & -2.66187629  & -1.68983501\\
                  &                 & -1.33267928  & -1.13394640\\
                  &                 & -1.03860624  & -1.00258200\\
                  \midrule
 $10^{-3}$        &  0.02613437    & -2.71516265  & -1.65622442\\
                  &                 & -1.32865305  & -1.12520220\\
                  &                 & -1.03745282  & -1.00045248\\
\bottomrule
\end{tabular}
}
\end{table}

\begin{figure}[h]
    \begin{center}
       \includegraphics[width=0.8\textwidth]{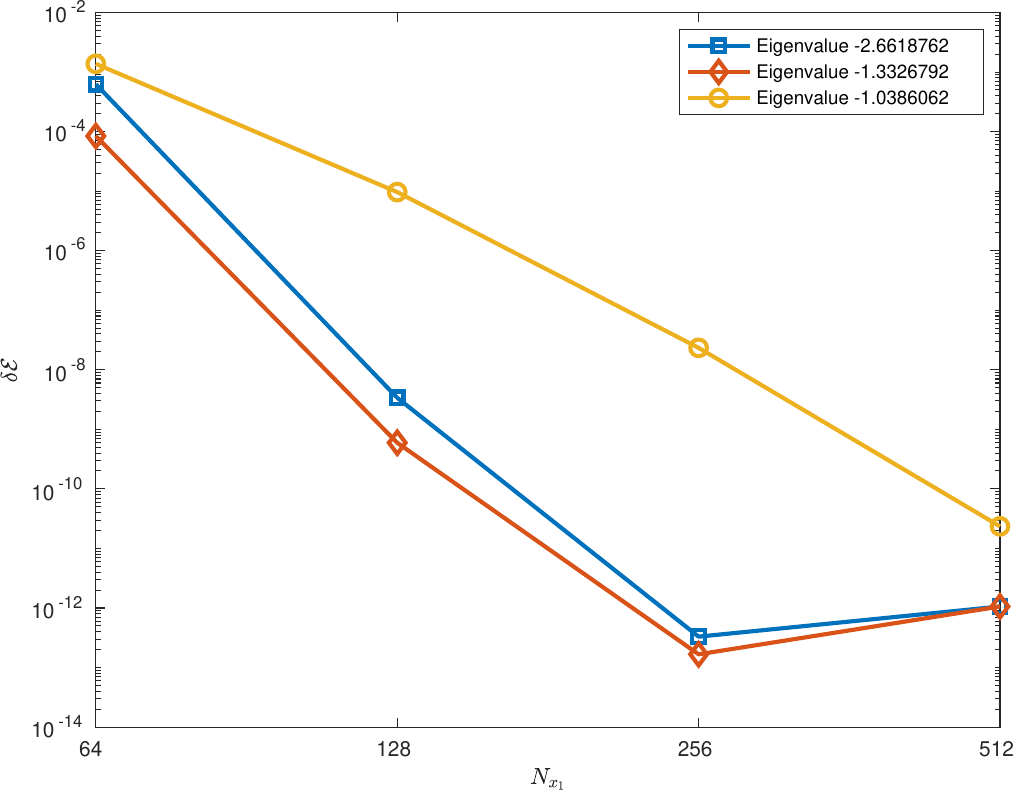}
    \end{center}
    \caption{Relative spectral discretization error $\delta\mathcal{E}$, Eq. \eqref{eq:error}, as the function of the grid size $N_{x_1}$ with $N_{y_1}=N_{x_1}/2$ for the three bosonic eigenvalues and the two-body binding energy $\mathcal{E}_0^{(2)}=10^{-2}$. The tolerance in the eigenvalue solver has been set to $10^{-12}$. 
    \label{fig:spectral}
    }
\end{figure}

In order to analyze the performance of our numerical scheme, we use a sequence of grid sizes $N_{x_1}$ and always choose $N_{y_1} = N_{x_1}/2$. 
Figure \ref{fig:spectral} shows the excellent convergence properties of the discretization. Indeed, the relative spatial discretization error
\begin{equation}
\label{eq:error}
\delta\mathcal{E}\equiv\frac{|\mathcal{E}(N_{x_1} \times N_{y_1})-\mathcal{E}(2N_{x_1}\times 2N_{y_1})|}{|\mathcal{E}(N_{x_1}\times N_{y_1})|},
\end{equation}
estimated as the relative difference of the computed eigenvalues on successive grids, is reduced exponentially until it reaches the tolerance of $10^{-12}$, which has been set in the solver for the computation of each eigenpair.

Next, we compare the convergence behavior and running time of three methods, namely the Krylov-Schur (KS) and the JDQR method without (no-prec) and with (prec) preconditioning, as described in Section~\ref{sec:precon}. In each case we exploit the tensor structure of the problem when applying the linear operator. 

For different two-body binding energies $\mathcal{E}_0^{(2)}$, we present in Table \ref{tab:summarizedresults1D} a comparison of the number of iterations and matrix-vector multiplications (MVM) with KS and no-prec/prec-JDQR methods necessary for computing the lowest three three-body energies in 1D with bosonic heavy particles, as listed in Table \ref{tab:2b_binding_energies}. All solvers have been set to the same tolerance $10^{-12}$. In addition, we have found similar behavior of these results in the case of fermionic heavy particles.

\begin{table}[h]
\caption{The number of iterations and matrix-vector multiplications (MVM) necessary to compute three bosonic eigenvalues, listed in Table \ref{tab:2b_binding_energies}, by using Krylov-Schur (KS) and Jacobi-Davidson methods, without preconditioner (no-prec-JDQR) and with preconditioner (prec-JDQR), respectively.
\label{tab:summarizedresults1D}}
\begin{tabular}{crrrrrrrr}
\toprule
$\mathcal{E}_0^{(2)} $	& $N_{x}$	&$ N_{y}$	&  KS	& MVM 	& no-prec-JDQR & MVM 	& prec-JDQR  & MVM \\
\midrule
$10^{-1}$	& 64 &	32	 & 134   & 2\,082	    & 56	& 1\,182   & 24   & 253\\
        	& 128 &	64	 & 1\,086	 & 16\,491	& 132	& 2\,920   & 28   & 258\\
           	& 256 &	128  & 5\,211	 & 80\,523	& 299	& 6\,782   & 28   & 261\\
        	& 512 &	256	 & 11\,324 & 172\,018	& 679	& 15\,526  & 37   & 506\\
        	& 1\,024 & 512 &       &          &       &        & 42   & 540\\
\midrule
$10^{-2}$	& 64 &	32	 & 88	 & 1\,390   & 46	& 932 	  & 25	  & 206\\
         	& 128 &	64	 & 720	 & 10\,999  & 92	& 2\,004    & 28 	  & 292\\
         	& 256 &	128	 & 6\,072	 & 91\,977  & 212	& 4\,765    & 29    & 337\\
        	& 512 &	256	 & 23\,546 & 357\,133 & 554	& 12\,648   & 30	  & 402\\
        	& 1\,024 & 512 &       &       &       &        & 39    & 502\\
\midrule
$10^{-3}$	& 64 &	32	 & 34   	& 560	   & 40	    & 792      & 23	 & 163\\
  	        & 128 &	64	 & 170   	& 2\,728	   & 74	    & 1\,596     & 25	 & 202\\
         	& 256 &	128	 & 2\,200    	& 33\,815	   & 148	& 3\,303     & 28	 & 235\\
         	& 512 &	256	 & 13\,599	& 207\,748   & 367	& 8\,342     & 28	 & 265\\
         	& 1\,024 & 512 &          &          &        &          & 34  & 271\\
\bottomrule
\end{tabular}

\end{table}

\begin{figure}
    \begin{center}
       \includegraphics[width=0.8\textwidth]{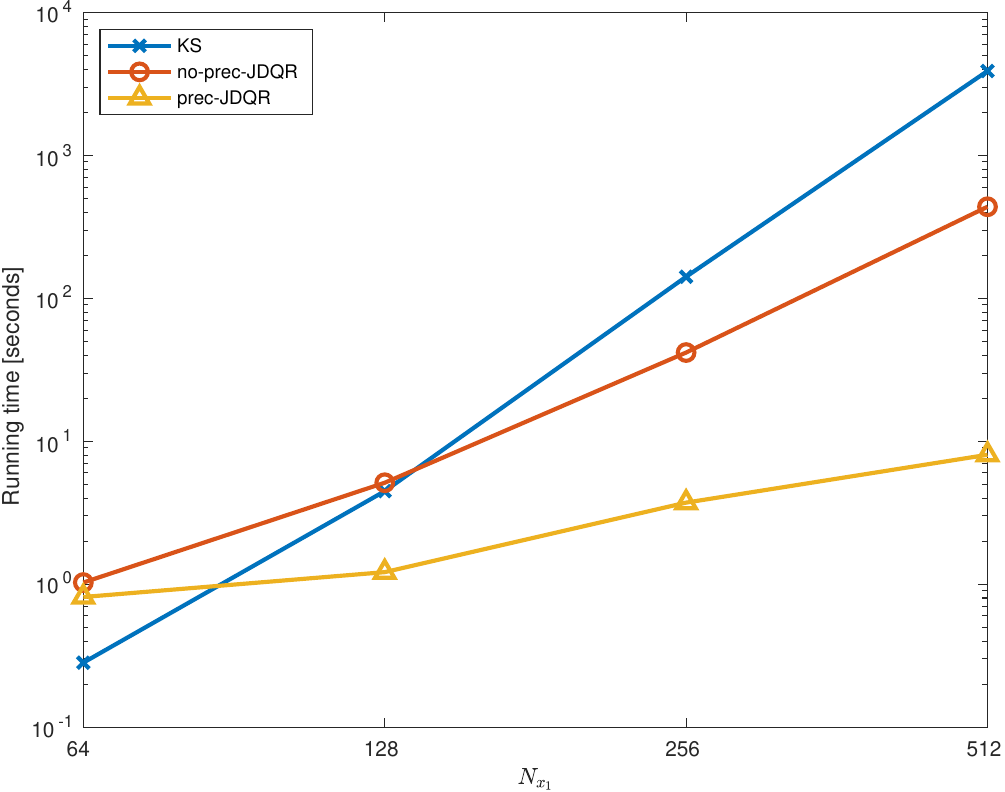}
    \end{center}
    \caption{The running time as the function of the grid size $N_{x_1}$, required for computing three bosonic three-body energies in 1D, Table \ref{tab:2b_binding_energies}, with the two-body binding energy $\mathcal{E}_0^{(2)} = 10^{-2}$, by using KS, no-prec-JDQR, and prec-JDQR methods. 
    \label{fig:runnungtime}
    }
\end{figure}

Compared to the KS method, the no-prec-JDQR one substantially reduces the number of matrix-vector multiplications, each of which is in fact an operator application of the discretized Hamiltonian, as described in Section~\ref{sec:op1D}. 
This is a consequence of the superior convergence rate of the inexact Newton process within JDQR over the Krylov subspace iteration.
The prec-JDQR method achieves an even more drastic reduction of the number of iterations and MVMs.
This results from the fact that the correction equation can now be solved to sufficient accuracy in order to achieve locally quadratic convergence of the Newton process.
In addition to MVM, this method requires a similar number of preconditioner applications, which have a similar cost, as discussed in Section~\ref{sec:precon}. Due to the fast convergence, the running time required for finding the eigenvalues is significantly reduced, as shown in Fig.~\ref{fig:runnungtime}. To perform this analysis, we have made use of a MATLAB implementation. In particular, the KS method has been realized via the MATLAB command \texttt{eigs}.

\subsection{Results for the three-body problem in 2D\label{sec:results2D}}

The aim of this section is to show the performance and viability of our approach in analyzing the three-body problem in two dimensions, as introduced in Section~\ref{sec:few-body_systems} and displayed in Fig.~\ref{fig:3body_scheme} (b). By choosing potentials of two very different shapes $f_\mathrm{G}$, Eq.~\eqref{eq:Gaussian_shaped_potential}, and $f_\mathrm{L}$, Eq.~\eqref{eq:Lorentzian_shaped_potential}, we find numerical evidence for a universal behavior in this system. In particular, we show that when the two-body binding energy $\mathcal{E}_0^{(2)}$ is decreased, the corresponding three-body energies for both interactions become more and more similar and approach those provided by a contact interaction between heavy and light particles. 

Our discretization scheme uses $N_x=N_{x_1}=N_{x_2}$ points in the $x_1$- and $x_2$-direction, and $N_y=N_{y_1}=N_{y_2}$ points in the $y_1$- and $y_2$-direction, leading to a total problem size of $N=N_x^2N_y^2$, where we choose again $N_y=N_x/2$.
To compute the three-body binding energies for the 2D case, we follow a similar procedure as presented in Section~\ref{sec:Results_1D}. 

First, for each two-body binding energies $\mathcal{E}_0^{(2)}=10^{-1}, 10^{-2},$ and $10^{-3}$, we determine the depth $v_0$ for the potential of Gaussian shape $f_\mathrm{G}$, Eq.~\eqref{eq:Gaussian_shaped_potential}, as well as of Lorentzian-cube shape $f_\mathrm{L}$, Eq.~\eqref{eq:Lorentzian_shaped_potential}, by solving the generalized eigenvalue problem~\eqref{eq:Schroedinger_equation_two_body} in 2D. The corresponding values of $v_0$ are listed in Table~\ref{tab:3b_binding_energies_2D}.

\begin{table}
\caption{\label{tab:3b_binding_energies_2D} Computed ratio $\mathcal{E}/\mathcal{E}_0^{(2)}$ of the three-body bound state in 2D for the Gaussian-shaped $f_\mathrm{G}$, Eq.~\eqref{eq:Gaussian_shaped_potential}, and the Lorentzian cube-shaped $f_\mathrm{L}$, Eq.~\eqref{eq:Lorentzian_shaped_potential} potentials and successively refined grids. The calculations are performed for bosonic identical particles and the mass ratio $\alpha=1$.\textbf{}}
\centering
\begin{tabular}{cccccc}
\toprule
  $\mathcal{E}_0^{(2)}$    & $f$ & $v_0$ in 2D & $(256\times 128)^2$ & $(440\times 220)^2$ & $(480\times 240)^2$ \\
\midrule
$10^{-1}$ & $f_\mathrm{G}$& 0.94734392 & -2.19995777 & -2.19995777  & -2.19995777\\
& $f_\mathrm{L}$& 1.64282612 & -2.22611678 &
-2.22611678&  -2.22611678  \\ \hline
$10^{-2}$ & $f_\mathrm{G}$& 0.48272728 &  -2.31159530 &  -2.31159833  &  -2.31159833 \\
& $f_\mathrm{L}$& 0.89384635 & -2.32245890   &    -2.32324354& -2.32324413   \\ \hline
$10^{-3}$ & $f_\mathrm{G}$& 0.31340752 &  -2.37039003  & -2.36675870   &  -2.36765166 \\
& $f_\mathrm{L}$& 0.59682960& -1.52944449  &    -2.36470275&   -2.37106085  \\ 

\bottomrule
\end{tabular}
\end{table}
 
Next, we use these parameters to solve numerically the corresponding three-body problem \eqref{eq:Schroedinger_equation_three_body} for the mass ratio $\alpha=1$. In this case there is only one eigenvalue $\mathcal{E}$ with $\mathcal{E}<-\mathcal{E}_0^{(2)}$. For each potential $f$ we use an increasing number of grid points to verify the numerical convergence of our results. Moreover, for a given grid size, we consecutively solve the eigenvalue problem for the three
values of $\mathcal{E}_{0}^{(2)}$ in decreasing order, using the previously computed subspace to start the next Jacobi-Davidson process (homotopy method). The corresponding energies of the single three-body bound state are listed in Table~\ref{tab:3b_binding_energies_2D}. 
 
For both potentials, the computed values of the three-body binding energies, relative to $\mathcal{E}_0^{(2)}$, indicate convergence for $\mathcal{E}_0^{(2)}=10^{-1}$ to at least nine digits. By further approaching the resonance, that is reducing the value of  $\mathcal{E}_0^{(2)}$, the convergence gets worse such that for $\mathcal{E}_0^{(2)}=10^{-3}$ and a grid size of $(480\times 240)^2$ only the first three digits of the ratio $\mathcal{E}/\mathcal{E}_0^{(2)}$ can be considered as converged.
 
Finally, we compare the converged values of the three-body binding energies for the grid size $(480\times 240)^2$ and different binding potentials of Gaussian shape $f_\mathrm{G}$ and Lorentzian-cube shape $f_\mathrm{L}$. Already for a two-body binding energy $\mathcal{E}_0^{(2)}=10^{-1}$, the three-body binding energies $\mathcal{E}=-2.199\ldots\mathcal{E}_0^{(2)}$ and $\mathcal{E}=-2.226\ldots\mathcal{E}_0^{(2)}$ are of the same order. By reducing the value of  $\mathcal{E}_0^{(2)}$ and moving closer to the resonance, the three-body energies further approach each other.

For the limiting case of a contact interaction between non-identical particles with $M=m$ in 2D, see Fig.~\ref{fig:3body_scheme} (b), a single three-body bound state has been predicted to exist  with an energy of approximately $2.39\ \mathcal{E}_0^{(2)}$~\cite{Brodsky2006}, $2.36\ \mathcal{E}_0^{(2)}$~\cite{Pricoupenko2010,Bellotti2011}, or $2.3896\ \mathcal{E}_0^{(2)}$~\cite{Petrov2020}. This result is only valid for identical bosonic  particles, as the same three-body system with identical fermionic ones does not have any bound state \cite{Brodsky2006,Pricoupenko2010}. Our numerical studies, which are summarized in Table~\ref{tab:3b_binding_energies_2D} and performed for two different local potentials of finite range, support these results. Thus, as  one of the central results of this article, we have shown that the considered three-body system in 2D displays a universal behavior as $\mathcal{E}_0^{(2)}\to 0$, that is the three-body states are independent of the details of the two-body interaction.

In the following, we provide additional details on the software implementation.
For our analysis, we have developed a hybrid MPI/multi-threaded C++ implementation to enable an efficient solution of the three-body problem.
Multi-threaded dense matrix products are provided by the Intel MKL (version 2020.4.304) and the JDQR method is implemented by phist~\cite{Thies2020phist}, version 1.9.6. 
The backend used within phist is the Trilinos library Tpetra~\cite{trilinos}, version 13.0.1.
In order to parallelize the application of the Hamiltonian, Eqs. \eqref{eq:discretized_Schroedinger_equation_1D} and \eqref{eq:Ham2D}, for the three-body problem, we use a column-wise distribution of
$W$ from Eq.~\eqref{eq:tensor_op2D} among the MPI processes, while the dense matrices, such as $\boldsymbol{D}_{x_1x_1}$ etc., constituting the Hamiltonian are replicated on all  processes. The communication involved in transposing the tensors $W$ and $\boldsymbol{D}_{xx}W^T$, Section \ref{sec:op2D}, can be overlapped with computations as follows:
\begin{enumerate}
\item Transpose the local columns of $W$;
\item For each column of $W^T$, dispatch a non-blocking `gather' operation;
\item Whenever a gather operation is finished for a local column of $W^T$, apply $\boldsymbol{D}_{xx}$ to that column;
\item The back transpose is then overlapped with the computation of the first term in Eq.~\eqref{eq:tensor_op2D}, $\boldsymbol{D}_{yy}W$.
\end{enumerate}

The numerical experiments were performed on the DelftBlue supercomputer at TU Delft with up to 220 compute nodes. Each node consists of two Intel Xeon E5-6248R processors with 24 cores and has 192 GB of RAM. On such a node the $(256\times 128)^2$ problem can be run by requiring about 8 GB per vector. In the absence of a preconditioner, we have used 8 nodes to accelerate the computation. The finest grid requires about 100 GB per vector. Its run on the full Phase 1 of DelftBlue (220 nodes) took about 3.5 hours for the three consecutive values of $\mathcal{E}_0^{(2)}$. A detailed performance analysis of the implementation and possible additional optimizations are beyond the scope of this article, but the observed running times in 1D, shown in Fig. \ref{fig:runnungtime}, give an indication of the cost of such simulations.

\section{Conclusion and outlook}
\label{sec:conclusion}

In this article, we present a novel, computationally-efficient tensor method to analyze the quantum-mechanical three-body problem with local two-body interactions in 1D and 2D. To build a matrix representation of the Schr\"odinger equation for the three-body problem, we have applied a pseudo-spectral method based on the rational Chebyshev polynomials. For the computation of the corresponding three-body binding energies, we have investigated different iterative methods for the diagonalization of the Hamiltonian matrix, namely the Krylov-Schur and the Jacobi-Davidson QR method with and without preconditioning. As a crucial point in implementing these methods, we have exploited the tensor product structure of the Hamiltonian to avoid storing redundant blocks of the matrix. Based on the direct solution of a Sylvester equation, we have developed an effective preconditioning strategy in the 1D case for accelerating a Jacobi-Davidson QR iterative eigensolver. In this way, the improved hardware efficiency of our tensor-based implementation has delivered a speed-up of about a factor 100 compared to the sparse matrix representation and Krylov methods that were utilized in previous studies of the three-body problem.

By developing a high performance implementation of our solution techniques that can be used on current supercomputers, we have shown for the first time the universal behavior of the 2D heavy-heavy-light three-body system when the ground-state energy of the heavy-light subsystems approaches zero. For this purpose, we have compared the numerically calculated three-body energies for different two-body interaction potentials of finite range. Close to the resonance they are approximately equal and coincide with the predicted three-body energy for a two-body contact interaction in the case of non-identical bosonic particles with equal masses. Thus, our newly developed tensor method is of crucial relevance for subsequent studies of the quantum-mechanical three-body problem in one and two spatial dimensions. 

As next steps, the methods developed in this article can straightforwardly be extended for studying other states of the three-body problems in 1D and 2D, such as virtual and resonant ones, as well as bound states embedded into the continuum. Moreover, since universality in three-body systems conventionally occurs in the region of very small binding energies, large grid sizes are necessary to obtain the required convergence. 
One way forward may be to enforce a low-rank structure on the occurring vectors. This approach would reduce the memory requirement for vectors in a similar way that we have used to reduce the memory requirement for the operator in this article.
Alternatively, by using polar coordinates we may exploit the conservation of the total angular momentum in the three-body system and thus reduce the effective dimension of the problem. However, the tensor product structure becomes more complicated in this case and parts of our implementation would require further developments.

In summary, we are convinced that our newly-developed tensor method will provide a deeper insight into the fascinating phenomena that occur for few-body problems in low dimensions.

\bibliographystyle{unsrt}
\bibliography{qwaves-tensor-paper}

\end{document}